\documentclass{SciPost}

\hypersetup{
    colorlinks,
    linkcolor={red!50!black},
    citecolor={blue!50!black},
    urlcolor={blue!80!black}
}

\usepackage[bitstream-charter]{mathdesign}
\usepackage{xspace} 
\usepackage{braket} 
\usepackage{changes} 

\DeclareSymbolFont{usualmathcal}{OMS}{cmsy}{m}{n}
\DeclareSymbolFontAlphabet{\mathcal}{usualmathcal}

\fancypagestyle{SPstyle}{
\fancyhf{}
\lhead{\colorbox{scipostblue}{\bf \color{white} ~SciPost Physics Core }}
\rhead{{\bf \color{scipostdeepblue} ~Submission }}

\fancyfoot[C]{\textbf{\thepage}}
}

\newcommand{\pcst}{$\textrm{pcst}^{\texttt{++}}$\xspace}
\newcommand{\pdagger}{{\phantom{\dagger}}} 

\newcommand{\Ham}{\mathcal H}
\newcommand{\Hamloc}{\mathcal H_\mathrm{loc}}
\newcommand{\HamM}{\Ham_\mathrm{matter}}
\newcommand{\HamL}{\Ham_\mathrm{light}}
\newcommand{\HamLM}{\Ham_\mathrm{LM}}

\newcommand{\refwithlabel}[2][]{%
    \hyperref[#2]{\ref*{#2}#1}}

\definechangesauthor[name=Kai, color=scipostblue]{KPS}
\definechangesauthor[name=Andi, color=scipostdeepblue]{AS}

\begin{document}

\pagestyle{SPstyle}

\begin{center}{\Large \textbf{\color{scipostdeepblue}{
To infinity and back -- $1/N$ graph expansions of\\light-matter systems
}}}\end{center}

\begin{center}\textbf{
Andreas Schellenberger\textsuperscript{$\star$} and
Kai P. Schmidt\textsuperscript{$\dagger$}
}\end{center}

\begin{center}
Friedrich-Alexander-Universität Erlangen-Nürnberg (FAU), Department of Physics, Staudtstraße 7, 91058 Erlangen, Germany
\\[\baselineskip]
$\star$ \href{mailto:andreas.schellenberger@fau.de}{\small andreas.schellenberger@fau.de}\,,\quad
$\dagger$ \href{mailto:kai.phillip.schmidt@fau.de}{\small kai.phillip.schmidt@fau.de}
\end{center}

\section*{\color{scipostdeepblue}{Abstract}}
\textbf{\boldmath{%
We present a method for performing a full graph expansion for light-matter systems, utilizing the linked-cluster theorem. This method enables us to explore $1/N$ corrections to the thermodynamic limit $N\to \infty$ in the number of particles, giving us access to the mesoscopic regime. While this regime is yet largely unexplored due to the challenges of studying it with established approaches, it incorporates intriguing features, such as entanglement between light and matter that vanishes in the thermodynamic limit.
As a representative application, we calculate physical quantities of the low-energy regime for the paradigmatic Dicke-Ising chain in the paramagnetic normal phase by accompanying the graph expansion with both exact diagonalization (NLCE) and perturbation theory (\pcst), benchmarking our approach against other techniques.
We investigate the ground-state energy density and photon density, showing a smooth transition from the microscopic to the macroscopic regime up to the thermodynamic limit.
Around the quantum critical point, we extract the $1/N$ corrections to the ground-state energy density to obtain the critical point and critical exponent using extrapolation techniques.
}}
\vspace{\baselineskip}

\noindent\textcolor{white!90!black}{%
\fbox{\parbox{0.975\linewidth}{%
\textcolor{white!40!black}{\begin{tabular}{lr}%
  \begin{minipage}{0.6\textwidth}%
    {\small Copyright attribution to authors. \newline
    This work is a submission to SciPost Physics Core. \newline
    License information to appear upon publication. \newline
    Publication information to appear upon publication.}
  \end{minipage} & \begin{minipage}{0.4\textwidth}
    {\small Received Date \newline Accepted Date \newline Published Date}%
  \end{minipage}
\end{tabular}}
}}
}


\vspace{10pt}
\noindent\rule{\textwidth}{1pt}
\tableofcontents
\noindent\rule{\textwidth}{1pt}
\vspace{10pt}


\section{Introduction}
\label{sec:intro}

Studying the collective interplay of matter with light is becoming a more and more promising route to alter properties of the constituents.
To achieve the needed strong coupling between light and matter, cavities in various forms are used to enhance the coupling between light and matter.
While the coupling of an individual particle with the light field is still small -- even within a cavity -- the coupling of a collective matter mode with light can alter physical properties like the low-energy spectrum or even induce quantum phase transitions \cite{Schlawin2019, Schuler2020, ParaviciniBagliani2018, Dmytruk2021, Dmytruk2024, Lenk2022,Kudlis2023,Sur2025}.

\begin{figure}[t]
    \centering
    \includegraphics{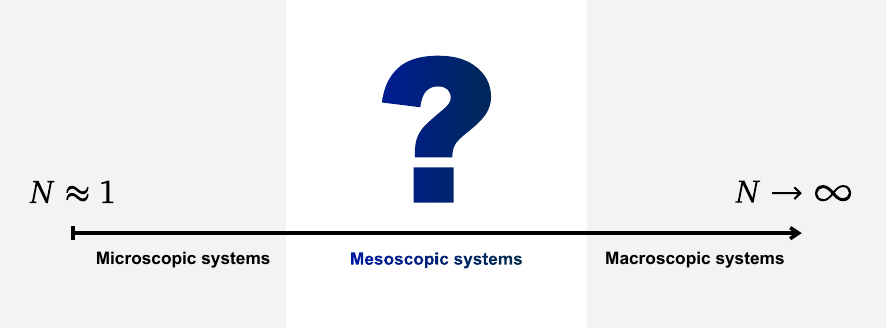}
    \caption{
        The three realms of light-matter quantum systems, depending on the number of matter entities $N$.
        While for the limiting cases of few particles and infinitely many a multitude of methods exist to model these systems, this is not the case for the intermediate mesoscopic regime.
        This regime is characterized by a large number of particles, such that both a few-particle approach and taking the thermodynamic limit are not valid.
        }
    \label{fig:mesoscopic}
\end{figure}

Modeling the typical setups normally starts from one of two limiting cases, considering the number of particles of the matter part \cite{Kudlis2023}, as visualized in Fig.~\ref{fig:mesoscopic}.
First, one can investigate the microscopic realm of a few particles coupled to the light mode.
In the limiting case of only one spin-1/2 particle and a single light mode, one ends up with the well-known Rabi model \cite{Rabi1936}.
Dealing only with a few particles can still make it feasible to search for analytical solutions or use numerical methods, like exact diagonalization, to describe the low-energy physics as the matter part can often be treated without truncation.
Second, one investigates the macroscopic domain in the thermodynamic limit, i.e., by taking the limit of the number of particles to be infinite.
This limiting case opens up the possibility of finding analytical solutions or simplifying numerical calculations \cite{RomanRoche2022,PerezSanchez2023,Schellenberger2024, Leibig2026,Sur2025}.

While both limits already offer a plethora of interesting physics emerging from the interplay of light and matter, current research indicates that there are missing aspects happening in-between \cite{Vidal2006,Lenk2022,Kudlis2023,Geng2025,DominguezNavarro2026,PerezSanchez2024,PerezSanchez2025,Barberena2025}.
This mesoscopic regime, where the number of particles is large but finite, such that the approximation of going to the thermodynamic limit is not valid, seems to host a plethora of stunning effects, which are only existing in this realm.
This reaches from altered scaling relations at critical points \cite{Vidal2006} to distinct dynamical effects reflected by observables \cite{Kudlis2023,Lenk2022,Geng2025,PerezSanchez2025, Barberena2025}.
Finding approaches to model this regime is both relatively new and tricky, as established methods from the limiting cases of small and large matter systems struggle with the exponential growth of the Hilbert space or approximations becoming less and less valid, respectively.

Inspired by graph expansions of classical and quantum many-body systems in the thermodynamic limit, we tackle the mesoscopic regime by performing a full graph expansion for a large class of light-matter models.
Graph expansions are an established tool for calculating quantities of interest of correlated matter systems, both perturbatively \cite{Yang2010,Yang2012,Powalski2013,Coester2015,Muehlhauser2024} and non-perturbatively \cite{Rigol2006,Kallin2013,Stoudenmire2014,Coester2015a,Hoermann2023}, depending on the graph solver that is used to perform calculations on the individual graphs.
The main benefit lies in the fact that the treatment of many graphs with comparably small Hilbert spaces can be used to obtain results in the thermodynamic limit by embedding these graphs in the infinite lattice \cite{Domb1974,OitmaaHamerZheng2006}.
In this work we generalize this approach to light-matter systems.
Because of the structure of these systems, we can extract a $1/N$ series expansion, yielding corrections to the thermodynamic limit.
In particular, with our approach we can access low-energy observables in the mesoscopic regime.

The article is organized as follows: In Sec.~\ref{sec:method} we will start by describing the method of linked-cluster expansions, first for pure matter systems, followed by the generalizations done to also treat light-matter systems.
The section will be concluded by discussing how to obtain the $1/N$ expansion, enabling us to study the mesoscopic regime.
In Sec.~\ref{sec:application} we will apply the method to the Dicke-Ising chain, being a paradigmatic model incorporating both light-matter and matter-matter interactions.
We will discuss our results in the mesoscopic regime and compare them to the limiting cases of small systems and the thermodynamic limit.
Finally, in Sec.~\ref{sec:conclusion} we summarize our findings and give a short outlook.

\section{Method}
\label{sec:method}

In this section we derive how to obtain a full linked-cluster expansion for light-matter systems.
For doing so, we first review the state-of-the-art way to calculate graph expansions for pure matter systems and introduce the typical setup.
Thereafter, we move on to light-matter systems, first focusing on the new ingredients -- namely a new type of graph -- that are added to the graph expansion.
Thus, while we focus physically on light-matter systems, the extensions introduced in this work can also be applied to other systems that incorporate this new type of graph, like bath modes coupling to many matter entities.
To treat this new graph type, we review the long-known case of disconnected clusters and adapt the approach to work with our light-matter setup.
We conclude the section by discussing how to obtain a $1/N$ expansion following the presented approach, accompanied with the limiting cases of the microscopic and macroscopic regime.

\subsection{Overview of linked-cluster expansions for matter systems}
\label{sec:linked-cluster-matter}

\begin{figure}
    \centering
    \includegraphics{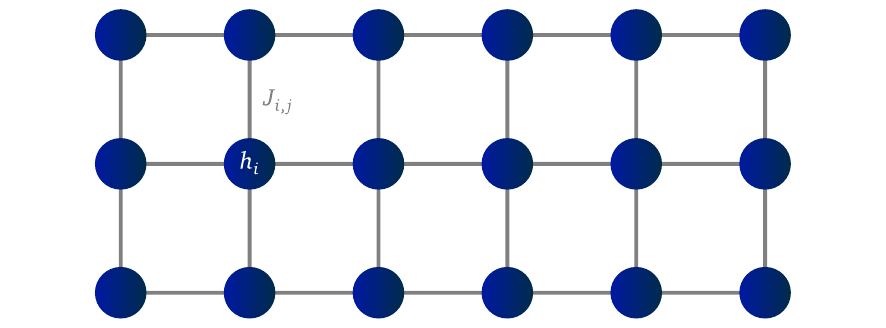}
    \caption{
        Visualization of an exemplary lattice structure defined by Eq.~\eqref{eq:ham}.
        The blue circles represent the matter degrees of freedom, which are coupled with their respective neighbors with $J_{i,j}$.}
    \label{fig:mlattice}
\end{figure}

First, we focus on linked-cluster expansions for pure matter systems. 
This approach is used for lattice-based quantum many-body models since some decades \cite{Domb1974, Gelfand2000, OitmaaHamerZheng2006,Knetter2000,Knetter2003_JoPA,Knetter2003}. 
For the following discussion, we define the class of Hamiltonians as
\begin{align}
    \Ham = \Hamloc + \mathcal V = \sum_i h_i O_i + \sum_{\braket{i,j}} J_{i,j} \,O_i O_j 
    \label{eq:ham}
\end{align}
with $O_i$ being a generic operator acting locally on site $i$ of a lattice $\mathcal L$ and $h_i,J_{i,j}$ some constants.
In Sec.~\ref{sec:application}, these operators will be spin operators realizing a magnetic field term and Ising interactions, respectively.
We denote $\Hamloc$ as the local and $\mathcal V$ as the non-local part of the Hamiltonian, assuming that the interaction terms in $\mathcal V$ are comparably small to the local terms in $\Hamloc$.
The structure of the lattice is defined by $\mathcal V$ where each term $J_{i,j}\,O_i O_j$ couples the two sites $i,j$ with each other.
While it is possible to also treat systems with long-range interactions \cite{Fey2019, Adelhardt2020, Adelhardt2025}, we will stick to the case that the sum only includes nearest-neighbor sites, denoted by $\braket{i,j}$.
We visualize an exemplary lattice structure in Fig.~\ref{fig:mlattice}.
We want to stress that the Hamiltonian could be generalized to host multi-site interaction terms, like done for the toric code in \cite{Muehlhauser2024,Kott2024}, not altering the following discussion.

As the Hamiltonian $\Hamloc$ is local, i.e., consists of a sum of operators only acting on one site of the lattice, it can be diagonalized trivially when the eigenbasis of the local operators $O_i$ is known.
Assuming that all $O_i$ have a unique ground state, we can directly induce a unique ground state for $\Hamloc$.
With this setup it is naturally possible to perform a linked-cluster expansion around the limit of $\Hamloc$ with $\mathcal V$ as a perturbation \cite{Gelfand2000}.
The driving idea for this approach is that we can calculate observables of interest of the low-energy subspace by only considering finite clusters in real-space, as the occurring quantum fluctuations caused by $\mathcal V$ are spatially confined to length scales of the order of the correlation length.
Such linked-cluster expansions can be set up both perturbatively \cite{Yang2010,Yang2012,Powalski2013,Coester2015,Muehlhauser2024} or non-perturbatively \cite{Rigol2006,Kallin2013,Stoudenmire2014,Coester2015a,Hoermann2023}, as we will discuss later in Subsubsec.~\ref{sec:graphsolvers}.

\subsubsection{Graph decomposition}
\label{sec:graph_decomposition}

In order to perform a graph expansion for a Hamiltonian [Eq.~\eqref{eq:ham}], one first has to fix which observable $\mathcal O$ of the low-energy subspace one aims to calculate.
When defining $\mathcal L$ as the lattice to consider, we denote $\mathcal O(\mathcal L)$ as the observable calculated on $\mathcal L$.
This is the quantity we are interested in.

It is usually not possible to determine $\mathcal O(\mathcal L)$ directly in the thermodynamic limit.
Instead, we take a finite cluster $C$ for calculation, chosen sufficiently large such that $\mathcal O(C) \approx \mathcal O(\mathcal L)$ for a non-perturbative treatment or $\mathcal O(C) = \mathcal O(\mathcal L)$ up to some finite perturbative order for a perturbative treatment.
Note that this is not a trivial statement but instead is motivated by the choice of the Hamiltonian as discussed before \cite{Domb1974,OitmaaHamerZheng2006}.

The idea of the graph expansion goes one step further.
Instead of directly calculating $\mathcal O$ on the large finite cluster $C$, we instead perform our calculations on subclusters of $C$.
Formally, we write
\begin{align}
    \mathcal O (C) = \sum_{c\subseteq C} o (c)
    \label{eq:graph-decomposition}
\end{align}
with a summation over all subclusters $c$ of $C$ (including $C$) and the reduced observable $o(c)$ on the subgraph
\begin{align}
    o(c) =  \mathcal O(c) - \sum_{c'\subsetneq c} o(c')\,,
    \label{eq:red-contr}
\end{align}
summing over the reduced contributions of all proper subclusters $c'$ over $c$ (excluding $c$) \cite{Domb1974,OitmaaHamerZheng2006}.
Intuitively, the reduced contribution of $c$ captures all fluctuations that can not be understood solely by those of the subclusters $c'$.
Investigating the sums over (proper) subclusters, one observes that many subclusters $c,\tilde c$ are isomorphic to each other, resulting in the same reduced contribution $o(c)=o(\tilde c)$.
Technically, we represent the sites of the (sub-)cluster as vertices and the bonds as edges of a representing graph.
The isomorphism of $c,\tilde c$ is then given as a color-preserving isomorphism of the representing graphs.
In the following, we will use the terms (sub-)graphs and (sub-)clusters equivalently, as we have established a suitable representation.
We can rewrite the sums in terms of distinct isomorphism classes, like
\begin{align}
    \mathcal O (C) = \sum_{\xi\subseteq C} N(\xi,C) \cdot o (\xi)\,,
    \label{eq:graph-decomposition-iso}
\end{align}
summing over all equivalence classes $\xi$ of subclusters of $C$ \cite{Domb1974,OitmaaHamerZheng2006}.
The embedding factor $N(\xi,C)$ denotes the cardinality of the equivalence class $\xi$, i.e., the number of times a subcluster of $\xi$ can be put on the cluster $C$.
For extensive observables $\mathcal O$ like the ground-state energy, the embedding factor $N(\xi,C)$ scales linearly with system size.
This is due to the translational invariance and the local interaction terms in the Hamiltonian $\Ham$.
This induces that $N(\xi,C)$ becomes a constant factor, when normalizing the extensive observable per site \cite{Domb1974,OitmaaHamerZheng2006}.
For calculating Eq.~\eqref{eq:graph-decomposition-iso} for a given observable $\mathcal O (C)$, we truncate the sum over all possible subclusters at some maximum size, e.g., the number of bonds in the respective graph.
In the following, we define the order of the graph expansion as the maximum number of bonds that is considered for the graphs in Eq.~\eqref{eq:graph-decomposition-iso}.

\subsubsection{(Non-)perturbative graph solvers}
\label{sec:graphsolvers}

What is left to do is calculating the observable $\mathcal O$ on the respective subgraphs $c$, i.e., calculating $\mathcal O(c)$ to obtain the reduced contribution $o(c)$ with Eq.~\eqref{eq:red-contr}.
The particular method to obtain $\mathcal O(c)$ is not determined by the graph decomposition in advance.
Instead, the choice of the `calculation method' depends on the particular insights one wants to extract from the study.
We can subdivide these approaches into the classes of non-perturbative \cite{Rigol2006,Kallin2013,Stoudenmire2014,Coester2015a,Hoermann2023} and perturbative \cite{Yang2010,Yang2012,Powalski2013,Coester2015,Muehlhauser2024} methods.
For this work, we will use one method of each type, namely exact diagonalization (ED) and \pcst \cite{Lenke2023}, which is an extension of the established perturbative continuous unitary transformation (pCUT) method, respectively \cite{Knetter2000,Knetter2003_JoPA, Knetter2003}.

With the ED method we have the option to calculate the contributions of the individual subgraphs with high numerical precision and are not restricted to a finite perturbative order.
While ED is limited to small clusters -- and thus in the approach of calculating $\mathcal O(C)$ directly also to small system sizes -- with the graph decomposition we can nonetheless extract the observable for larger finite-size systems or directly in the thermodynamic limit \cite{Rigol2006}.
Of course the quality of the results is closely connected to the convergence of the graph decomposition, i.e., how big of an error we make when only considering graphs up to some finite size in Eq.~\eqref{eq:graph-decomposition}, as discussed at the beginning of this section.
This length scale, defined by the maximum size of the graphs, is connected to the correlation length and therefore to the energy gaps in the system under study.
Another inherent limitation of this method is the need to fix all parameters of the Hamiltonian for the calculations.
While we can nonetheless do sweeps over parameter regimes of interest, we are not able to capture the direct correspondence of a certain parameter and the observable, as it is possible with other methods.
We will revisit this issue later, when discussing the extraction of the $1/N$ series expansion of observables in light-matter systems in Subsubsec.~\ref{sec:mesoscopic}.

Second, we use \pcst as a perturbative method \cite{Lenke2023} taking $\Hamloc$ as the unperturbed Hamiltonian and $\mathcal V$ as the perturbation. 
The \pcst is an extension of the established pCUT method  \cite{Knetter2000,Knetter2003_JoPA, Knetter2003}, which was already used for a variety of quantum many-body systems \cite{Knetter2000,Knetter2003_JoPA, Knetter2003,Fey2019,Muehlhauser2024,Kott2024,Adelhardt2025}.
While pCUT is limited to Hamiltonians with a single ladder spectrum in the unperturbed part $\Hamloc$, the \pcst method can also be applied to systems with any finite number of ladder spectra in $\Hamloc$ and systems with dissipative terms that can be described by Lindbladians \cite{Lenke2023}.
For closed systems this nonetheless restricts the realm of possible Hamiltonians to those that have a fitting spectrum for all $O_i$.
We will use the \pcst method for the combined light-matter system in Sec.~\ref{sec:application}, which has two ladder spectra, coming from the matter and light degrees of freedom, respectively.
Although there is no necessity to use it, such an perturbative method is an inherently fitting approach to be used with the graph expansion \cite{Yang2010}.
We can choose the local Hamiltonian $\Hamloc$ both as the starting point for the graph expansion and for the perturbation theory to ensure good convergence around this limit.
Thus, we obtain the observable $\mathcal O$ as a series over the perturbation parameters $J_{i,j}$.
The maximum perturbative order is directly linked to the order of the graph expansion, as introduced at the end of Subsubsec.~\ref{sec:graph_decomposition}.
The other way around this implies that we only have to calculate the graphs up to some specific order when being interested in the exact determination of a fixed maximum perturbative order.
The \pcst method can also preserve a general ratio between the spacings of the individual ladder spectra at the cost of longer computation times.
For more information on the \pcst method, see App.~\ref{sec:perturbative} and \cite{Lenke2023}.
In summary, while the perturbative ansatz is restricted to the phase imprinted by $\Hamloc$, we can derive analytical expressions, depending on the parameters of the Hamiltonian, for the observable of interest.

\subsection{$1/N$ linked-cluster expansions for light-matter systems}

The previous subsection was intended to give an introduction to the established approach of linked-cluster expansions of matter systems.
In the following subsection we will introduce what is new to the method discussed in this work.
Physically speaking, we are enlarging the space of treatable models from pure matter systems to those which also have light-matter couplings.
Technically speaking, we incorporate a new lattice structure that leads to a novel graph type that can be treated as disconnected matter graphs.
As non-linked-cluster expansions are also well established \cite{Domb1974,OitmaaHamerZheng2006}, we can use this knowledge to build up a linked-cluster expansion of light-matter systems.

\subsubsection{Novel graph type}

We start by defining a generalized setup that includes all systems we are interested in.
To make things clear we subdivide the Hamiltonian explicitly into the matter and light part:
\begin{align}
    \Ham &= \HamM + \HamL + \HamLM\,, \label{eq:generalized-hamiltonian}\\
    \HamM &= \sum_i h_i O_i + \sum_{\braket{i,j}} J_{i,j}\, O_i O_j\,,\\
    \HamL & = \sum_k \omega_k a^\dagger_k a_k^\pdagger\,, \label{eq:HamL}\\
    \HamLM & = \sum_k (a^\dagger_k + a_k^{\phantom{\dagger}}) \sum_{i} g_{k,i} O_i\,.
\end{align} 
While $\HamM$ is the known Hamiltonian from Eq.~\eqref{eq:ham}, $\HamL$ describes a finite number of light modes with frequencies $\omega_k$, with $a^\dagger_k, a_k^\pdagger$ being the bosonic creation and annihilation operators of the $k$-th light mode \cite{Walls2025}.
The third term $\HamLM$ describes the linear coupling between light and matter \cite{Walls2025}.
The crucial novel aspect is that the notion of neighborhood changes for $\HamLM$.
While we defined $\HamM$ to only have nearest-neighbor interactions, $\HamLM$ couples \emph{all} sites $i$ of the matter lattice to any light mode $k$.

\begin{figure}[t]
    \centering
    \includegraphics{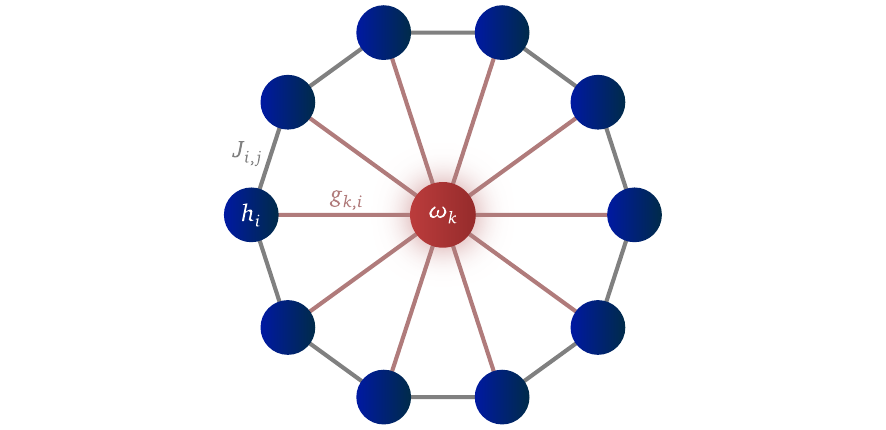}
    \caption{
        Visualization of an exemplary generalized lattice structure from Eq.~\eqref{eq:generalized-hamiltonian}.
        A chain of $N=10$ matter sites with periodic boundary conditions is coupled to a single light mode.
        The blue circles represent the matter degrees of freedom, the red circle represents the light mode.
        While the matter degrees of freedom are coupled only to their nearest neighbors with $J_{i,j}$, the light mode is coupled to all matter sites with $g_{k,i}$.}
    \label{fig:lmlattice}
\end{figure}
We sketch this new generalized lattice structure in Fig.~\ref{fig:lmlattice}, restricting ourselves to a single light mode.
We can define a combined light-matter graph, with representing the light mode by a site, which is connected to all matter sites.
To distinguish the different sites and bonds, we use colored graphs, as indicated in the figure.

From the perspective of graph theory, the Hamiltonian of Eq.~\eqref{eq:ham} already describes the light-matter systems of interest, when generalizing the scheme of nearest-neighbor sites, denoted by $\braket{i,j}$.
First, we can directly map $\HamL$ to additional sites with distinct local operators $O_i$ in Eq.~\eqref{eq:ham}.
The light-matter coupling can be incorporated by the second term of Eq.~\eqref{eq:ham} by defining the nearest-neighbor relation $\braket{i,j}$ in that way that any matter site is neighbor of any light-mode site, in accordance to Fig.~\ref{fig:lmlattice}.
While with this more compact Hamiltonian the physical origin stays more obscure, it unravels the fact that adding light to a pure matter system is mainly about changing the qualitative lattice structure, apart from distinct bosonic operators for the light modes.

\begin{figure}[t]
    \centering
    \includegraphics{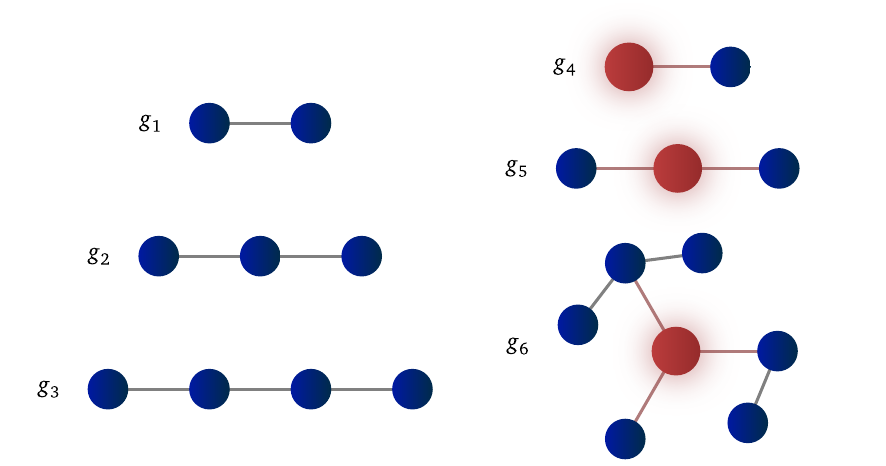}
    \caption{
        Visualization of some of the subgraphs of the lattice depicted in Fig.~\ref{fig:lmlattice}. 
        On the left side, graphs $g_1,g_2,g_3$ represent pure matter graphs. 
        These have embedding factors that scale linearly with system size and are treated with standard graph-decomposition techniques described in the last section. 
        On the right side, graphs $g_4,g_5,g_6$ represent the new type of graph, stemming from the light-matter interactions. 
        In general, these graphs have embedding factors that scale polynomially in system size. 
        This is grounded in the lattice structure of Fig.~\ref{fig:lmlattice}, e.g., the light mode is connected to all matter sites.}
    \label{fig:lmgraphs}
\end{figure}

Coming back to the foundational Eq.~\eqref{eq:graph-decomposition} for the graph decomposition, we now have to examine which new subgraphs $c$ occur.
While the sum still contains all pure matter graphs, we additionally have to consider also graphs with light-matter interactions, e.g., graphs with a bond formed by this interaction.
We visualize some of these new graphs (along with the matter-only graphs) in Fig.~\ref{fig:lmgraphs} for the lattice in Fig.~\ref{fig:lmlattice}.
According to Eq.~\eqref{eq:graph-decomposition-iso}, we have to determine both the embedding factor and the reduced observable of the new graphs.
The calculation of the reduced observables is a quite straight-forward generalization of the calculations for pure matter systems, as we solely have to obtain the observable of interest on the given local cluster.

The qualitative new aspect is the embedding factor of the light-matter graphs.
While the embedding factor for matter systems with short-range interactions scales linearly with system size \cite{Domb1974,OitmaaHamerZheng2006}, the embedding factor for the light-matter graphs has to be expressed by a polynomial in system size.
Exemplarily, this can be seen for the subgraph $g_5$ in Fig.~\ref{fig:lmgraphs}.
Each of the matter sites can be placed on all sites of the lattice, independently, except for the hard-core constraint.
This gives an embedding factor of $N(N-1)/2$ with $N$ being the number of matter sites.
The factor of $1/2$ comes from the indistinguishability of the two matter sites \cite{Domb1974}.

While the calculation of the embedding factor for such simple subgraphs and lattices is straightforward, we have to establish a general scheme for arbitrary lattices of the type defined above.
This is what we will discuss next. 

\subsubsection{Treatment of disconnected matter clusters}

The new class of graphs, introduced in the last subsubsection, connects matter sites of arbitrary distance mediated by the light modes.
This independence of the matter geometry leads to an embedding factor that can scale not only linear but also in higher orders of the system size.

\begin{figure}[t]
    \centering
    \includegraphics{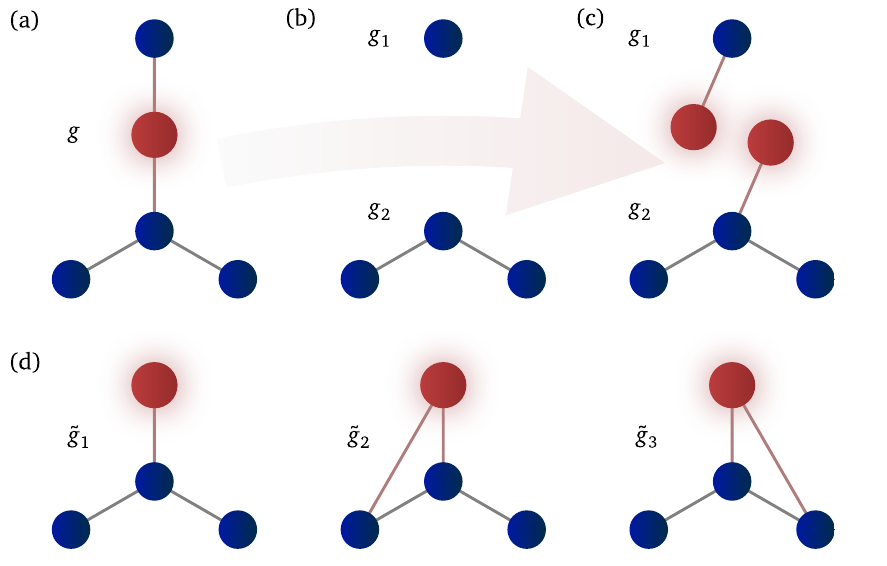}
    \caption{
        Sketch of treating the light-matter graphs as disconnected pure matter graphs.
        (a): The starting graph $g$ couples two independent matter graphs via the light-matter coupling.
        (b): When ignoring the light modes and their light-matter interactions, we end up with pure matter graphs.
        For this exemplary graph, we have two disconnected graphs $g_1,g_2$. 
        For calculating the embedding factor of this pure matter graph, we can use Eq.~\eqref{eq:embedding-disconnected}, as explained in the main text.
        (c): To take the light-matter interaction back into account, we insert them back for every disconnected graph $g_1,g_2$ individually. With these modified disconnected graphs, we can still use Eq.~\eqref{eq:embedding-disconnected} to calculate the embedding factor, enforcing that the two light sites have to overlap.
        (d): Overlap partitions $\tilde g_1, \tilde g_2, \tilde g_3$ forming out of the graphs $g_1, g_2$ from (c).
        The graphs $\tilde g_2, \tilde g_3$ are isomorphic.
        Therefore, we only have to take into account one of the two.
        }
    \label{fig:disconnectedgraphs}
\end{figure}

When ignoring the mediating light modes and their light-matter interactions for a second, we are left with disconnected pure matter graphs, that may have some local structure induced by $\HamM$, as depicted in Fig.~\ref{fig:disconnectedgraphs}.
With this ansatz, the problem of calculating the embedding factor of this new type of graphs boils down to the problem of calculating the embedding factor of these disconnected graphs on the matter lattice.
This task was already solved some decades ago when performing high-temperature expansions \cite{Domb1974, OitmaaHamerZheng2006}.
In the following, we will sketch the approach for these disconnected graphs, following \cite{Domb1974}.
Note, we still have to deal with our ad hoc step to ignore the light modes and their light-matter interactions, as this will require a slight adaption of the algorithm, described in the following.

To calculate the embedding factor of a disconnected graph $g$ on a larger graph $G\supseteq g$, we assume that we know already the embedding factor of all connected subgraphs of $g$.
We can furthermore claim that $g=g_1 \cup g_2$, with $g_1\neq g_2$ being subgraphs and $g_1\cap g_2 = \emptyset$.
If $g$ would consist out of more than two disconnected subgraphs, we can iterate over pairs of these subgraphs to obtain the embedding factor of $g$ recursively.
One finds the embedding factor of $g$ to be
\begin{align}
    N(g,G) = N(g_1,G) \cdot N(g_2,G) - \sum_{\tilde g} \{g_1+g_2 = \tilde g\} \cdot N(\tilde g,G)\,,
    \label{eq:embedding-disconnected}
\end{align}
with a summation over all graphs $\tilde g$ whose number of sites is smaller than the sum of the sites of $g_1$ and $g_2$ and that are a subgraph of $G$.
The expression $\{g_1+g_2 = \tilde g\}$ is defined as the number of overlap partitions $\tilde g$ into $g_1,g_2$.
Loosely speaking we therefore count the number of ways we can arrange $g_1,g_2$ (with a potential overlap) such that the union is isomorphic to $\tilde g$.
For more details, see \cite{Domb1974}.

Examining Eq.~\eqref{eq:embedding-disconnected}, we first multiply the embedding factors of the subgraphs $g_1,g_2$.
This is a reasonable quantity, thinking about the simple example of $g_5$ in Fig.~\ref{fig:lmgraphs}, where we placed the two matter sites almost independently from each other onto the lattice.
The correction to this simple approach is done with the sum over $\tilde g$.
Here, all forbidden overlaps of $g_1,g_2$ are subtracted again that we wrongfully counted when we multiplied the embedding factors of the subgraphs.

Last, a comment on the case $g_1=g_2$. For this situation, we have to adjust Eq.~\eqref{eq:embedding-disconnected} slightly to avoid double counting caused by the fact that $g_1,g_2$ are indistinguishable \cite{Domb1974}.

So, we reviewed the approach to calculate the embedding factors of disconnected graphs.
It remains to connect this ansatz to the light-matter setting, which we got rid of by removing light modes and light-matter interactions.
When taking the nomenclature from above, we adapt the approach of subdividing $g$ into $g_1,g_2$, by adding the respective bonds and sites from $\HamL, \HamLM$ to $g_1,g_2$, individually, as visualized in Fig.~\ref{fig:disconnectedgraphs}.
Thus, we end up with the subgraphs' intersection $g_1\cap g_2$ consisting of a finite number of light sites.
Nevertheless, we can still use Eq.~\eqref{eq:embedding-disconnected} to calculate the embedding factor, as for the pure matter system, when enforcing the individual light sites of $g_1,g_2$ to overlap.
This can be ensured by using distinct colorings of the light sites for all light modes, as defined in Eq.~\eqref{eq:HamL}.
Keep in mind that the respective embedding factors of $g_1,g_2$ may differ from the previous pure-matter case, where we neglected the light part, as can be seen by the possible overlap partitions in Fig.~\refwithlabel[(d)]{fig:disconnectedgraphs}.

To sum things up, we can extract the embedding factor of graphs with light-matter bonds by treating the subgraphs, \emph{only} connected via the light sites, as disconnected graphs, using the established approach of high-temperature expansions.

\subsubsection{On different types of graph expansions}
\label{sec:typegraphexpansion}

Up to now, we concentrated on the most general form of a graph expansion, often called \textit{full graph expansion}.
For a full graph expansion we consider all possible subclusters $c$ of a given finite cluster $C$ for our calculations, according to Eq.~\eqref{eq:graph-decomposition}.
It turns out that for some models it is beneficial to not do a full graph expansion, but expand according to a `larger structure' of the system.
So, not all possible subgraphs are considered but only a closed subset, in the sense of all overlap graphs being included in the subset.
This prerequisite is needed to be able to calculate the reduced contributions of observables, as given in Eq.~\eqref{eq:red-contr}.
The general advantage of other types of graph expansions lies in the reduced number of graphs potentially enabling calculations up to higher order and therefore better convergence.
A well-known example are rectangular-graph expansions, which expand in terms of rectangles to treat models on the square lattice \cite{Enting1977,Kallin2013,Coester2015a, Ixert2016, Gan2020a}.

For our given light-matter models of Eq.~\eqref{eq:generalized-hamiltonian}, a natural type of expansion is a \textit{full light expansion}.
This type of expansion restricts to subclusters $c$ where all matter sites are connected to all given light modes.
Effectively, this corresponds to a matter-only graph expansion, where the light-matter interactions are ignored in the first place and are added later to every generated matter graph.
This brings the expansion even closer to the established non-linked cluster expansions, performing the same calculations on the pure matter part and only adding the light part as a final step.
We define the order of the expansion as the number of matter-matter bonds in the respective graph.
We motivate the full light expansion to be especially beneficial for parameter regions that are dominated by local and matter-matter interactions.
The added light-matter interactions thus act as a `correction' to the bare matter model.
We will compare the two types of expansion in Sec.~\ref{sec:application}.

As a note for completeness: We will use an additional expansion for the perturbative method, mentioned in Subsubsec.~\ref{sec:graphsolvers}, which we will denote as \textit{full matter expansion}.
Namely, we expand along the light-matter interactions, ignoring the matter-matter interactions in the first place.
In a final step, the matter-matter interactions are -- analogous to the full light expansion -- emplaced back to all graphs such that every matter site has all possible matter-matter interactions embedded.
This expansion is motivated by the concrete model investigated in Sec.~\ref{sec:application}, where we can solve the matter part exactly, as it is diagonal.
This enables us to only treat the light-matter interaction perturbatively, by taking the neighborhood of matter sites into account, which is non-negligible due to the Ising interaction.
Analogous to above, we define the order of the expansion as the number of light-matter bonds in the respective graph.

\subsubsection{Approaching the mesoscopic regime}
\label{sec:mesoscopic}

We are now able to calculate observables of interest $\mathcal O$ for light-matter Hamiltonians using a linked-cluster expansion.
With this last method subsubsection, we want to focus on the form of the results we get from the expansion.

As discussed in the last subsubsection, the embedding factors do not necessarily scale linearly with system size anymore but have to be written as a polynomial. 
We can thus write the embedding factor of a subgraph $g$ of the graph $G$ as
\begin{align}
    N(g,G) = \sum_{o=1}^\infty c_o N^o
\end{align}
with $N$ being the number of matter sites of the system and $c_o\in \mathbb Z$ the constants we calculate.
Assuming that $\mathcal O$ is an extensive quantity, i.e., scaling linearly with $N$, we can induce from Eq.~\eqref{eq:graph-decomposition-iso} that the reduced quantities have to scale in powers of $1/N$ such that higher orders in $N$ cancel.
This finding is in line with respective Hamiltonians, introduced in Eq.~\eqref{eq:generalized-hamiltonian}.
Because of the lattice structure, one has to introduce renormalization constants for the respective couplings to keep the Hamiltonian's observables, like energies, from scaling superextensively.
This can be either motivated physically by limited interaction volumes, yielding a $1/\sqrt N$ factor for $\HamLM$ \cite{Zhang2014, Svendsen2023}, or it can be introduced by rescaling the Hamiltonian, e.g., applying Kac's rescaling \cite{Kac1963, Campa2003, Mori2012,GonzalezLazo2021, Mattes2024, RomanRoche2023}.
This is also the case for the Dicke-Ising model \cite{Zhang2014, Rohn_2020}, which we will discuss in Sec.~\ref{sec:application}.

When dividing the extensive quantity $\mathcal O$ of Eq.~\eqref{eq:graph-decomposition-iso} by $N$, we obtain an intensive observable with finite-size corrections in form of a series expansion in $1/N$, namely
\begin{align}
    \mathcal O(G) / N = \sum_{o=0}^\infty d_o N^{-o}\,,
    \label{eq:finite-size-series-expansion}
\end{align}
with $d_o$ depending on the parameters of the Hamiltonian but not on $N$.
Note that we need to calculate the reduced quantities from Eq.~\eqref{eq:graph-decomposition-iso} as a series expansion to keep track of the \mbox{$N$-dependency}, as discussed in Subsubsec.~\ref{sec:graphsolvers}.
This is therefore a major benefit in contrast to numerical methods, like ED, with which we are not able to determine Eq.~\eqref{eq:finite-size-series-expansion}. 

With our approach we are therefore expanding the observable $\mathcal O$, starting from the thermodynamic limit $N\to\infty$ to finite systems.
To calculate higher orders in $o$, one has to include larger graphs, as the scaling $N^{-o}$ is connected to the light-matter bonds, as discussed in the paragraph above.
Therefore, the sum in Eq.~\eqref{eq:finite-size-series-expansion} will normally be truncated to finite $o$, inducing worse results for small $N$ in comparison to large $N$ close to the thermodynamic limit.
In addition, the quality of the series expansion also depends on the validity of the graph expansion around a local limit, as discussed in Subsec.~\ref{sec:linked-cluster-matter}.
Our general strategy is connected to other methods that derive $1/N$ expansions for molecular polaritonic systems with no direct matter-matter interaction but more complex matter structures.
Analogous to our approach, they find corrections to the thermodynamic limit, when going to systems with more `effective molecules' (like higher-order graphs) to incorporate finite-size effects \cite{PerezSanchez2023,PerezSanchez2024}. 

We obtain the observable in the thermodynamic limit $N\to\infty$ by neglecting all $o\neq 0$ in the sum of Eq.~\eqref{eq:finite-size-series-expansion}, yielding 
\begin{align}
    \lim_{N\to\infty}\mathcal O(G) / N = d_0\,.
\end{align}
When we are only interested in the thermodynamic limit, we can thus limit our graph expansion of Eq.~\eqref{eq:graph-decomposition-iso} to contributions that do not scale with $N$.

Putting things together, we introduced a graph-expansion method for light-matter systems by using the approach of disconnected matter graphs, known from high-temperature expansions \cite{Domb1974,OitmaaHamerZheng2006}.
By calculating quantities of interests in this way, we obtain a series expansion out of the thermodynamic limit with finite-size corrections.
We therefore have a handle to not only calculate observables in the thermodynamic limit but also to get insights of the mesoscopic regime with large finite numbers of particles.

\section{Application onto the Dicke-Ising model}
\label{sec:application}

After introducing the graph-expansion approach for light-matter systems, in this section we apply the method on the paradigmatic Dicke-Ising model \cite{Zhang2014}. 
While the graph-expansion method is applicable also in higher spatial dimensions, we focus on the one-dimensional case in the following.
First, we give an introduction to the Dicke-Ising model followed by the discussion of the obtained results.
We compare our findings, calculated with the graph-expansion ansatz, with other methods, namely ED for small systems and an effective theory for the infinite system \cite{Schellenberger2024}.

\subsection{Overview of the model}

The Dicke-Ising model can be understood as a minimal model that both includes light-matter interactions and direct matter-matter interactions.
The Hamiltonian is formed as a combination of the Dicke model \cite{Dicke1954,Hepp1973} with an Ising interaction between neighboring sites.
It is written as \cite{Zhang2014}

\begin{align}
    \mathcal H_\mathrm{DI} =  \frac {\omega_0} 2 \sum_j \sigma^z_j + \omega a^\dagger a + \frac{g}{\sqrt N} (a^\dagger + a ) \sum_j \sigma_j^x + J \sum_{\braket{j,l}} \sigma^z_j \sigma^z_l
    \label{Eq:DickeIsing}
\end{align}
with ferromagnetic (antiferromagnetic) Ising interactions for $J<0$ ($J>0$) and $N$ being the number of sites.
The Hamiltonian can be mapped to the general Hamiltonian Eq.~\eqref{eq:generalized-hamiltonian} of Sec.~\ref{sec:method}, restricting $\HamL$ to a single light mode at $k=0$ and introducing spin operators for the generic operators $O_j$.

Due to the two kinds of interactions between the individual entities, the model features a number of phases at zero temperature.
The phase diagram was studied by a number of works, with different methods, including mean-field theory \cite{Zhang2014}, ED \cite{Rohn_2020}, quantum Monte Carlo \cite{Langheld2025}, effective theories \cite{Schellenberger2024}, and variational approaches \cite{Leibig2026,Mendonca2025} for various dimensions.
To be noted, the results in \cite{Mendonca2025} differ qualitatively from the other studies.
In \cite{CommentsOnMendonca2025} it is demonstrated and argued how to find the features that are missing in \cite{Mendonca2025}.

\begin{figure}
    \centering
    \includegraphics{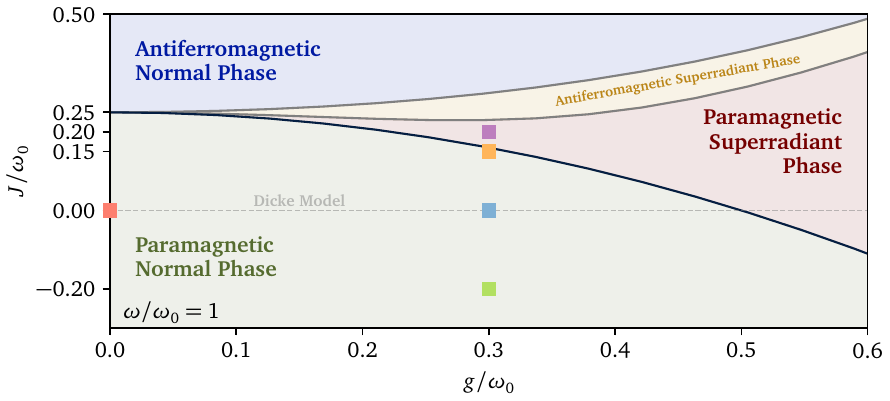}
    \caption{
        Phase diagram of the Dicke-Ising chain of Eq.~\eqref{Eq:DickeIsing}.
        The ratio of the magnetic field and the photon frequency is fixed to $\omega / \omega_0 = 1$.
        The phase transition lines obtained by mean-field theory \cite{Zhang2014} are plotted for orientation.
        The dark-colored line indicate a quantitative correct phase transition in the considered parameter regime, while the greyed-out lines indicate that mean-field theory deviates from quantitative methods \cite{Langheld2025, Leibig2026}.
        The data points we investigate in this work are plotted with different colors for easy recognition in the following figures.
        }
    \label{fig:phasediagramDI}
\end{figure}

Due to the various applied methods, including quantitative studies, it is settled that the model realizes four phases in all spatial dimensions as depicted in the phase diagram in Fig.~\ref{fig:phasediagramDI}.
The phases can be characterized by the photon density $\braket{a^\dagger a}/N$ and the $x$-magnetization of the spins $\sum_j \braket{\sigma_j^x}/N$.
Having small light-matter interactions, the matter-matter couplings dominate, realizing either the `paradigmatic normal phase' or the `antiferromagnetic normal phase'.
Going to larger light-matter interactions, one observes phase transitions to a `paradigmatic superradiant phase' -- also known from the pure Dicke model \cite{Dicke1954,Hepp1973,Emary2003a} -- and an intermediate `antiferromagnetic superradiant phase' where the matter part is still ordered by the Ising term.
For both superradiant phases the photon density is non-vanishing.

Most of the studies focus on the thermodynamic limit, either extrapolating their finite-size findings to the thermodynamic limit \cite{Langheld2025,Mendonca2025} or working directly in the thermodynamic limit with an effective model \cite{Zhang2014, Leibig2026, Schellenberger2024,RomanRoche2025}.
In contrast, the study by Vidal and Dusuel focuses on finite-size corrections for the pure Dicke model ($J=0$) with an analytical approach to determine scaling exponents at the critical point \cite{Vidal2006}.

As motivated in the methods section, the graph expansion is built up starting from the limit of a local  Hamiltonian.
For the case of the Dicke-Ising Hamiltonian this is the case when setting $g=J=0$.
The most natural start is therefore investigating the paradigmatic normal phase.
Nonetheless, one can also investigate other phases, keeping in mind that the convergence is potentially bad.
This can be cured by introducing edge terms and an expansion scheme that is fitted for the underlying dominant structure.
We refer to respective works investigating the Ising limit for pure matter systems, which could also be done for the Dicke-Ising model in future works to investigate the antiferromagnetic phases of the Dicke-Ising model \cite{Dusuel2010,Hoermann2023}, as done in \cite{Leibig2026} in the thermodynamic limit.

\subsection{Results in the mesoscopic regime in 1D}

In the following subsection we present and discuss our results for the one-dimensional Dicke-Ising model.
We set $\omega_0 = \omega = 1$ for Eq.~\eqref{Eq:DickeIsing}, leaving the light-matter and matter-matter interaction strength to be varied.
We treat $J,g$ as small compared to $\omega_0,\omega$ to stay within or close to the paramagnetic normal phase.
As motivated in the last section, this resembles the natural starting point for the graph expansion around local entities.

\begin{figure}[t]
    \centering
    \includegraphics{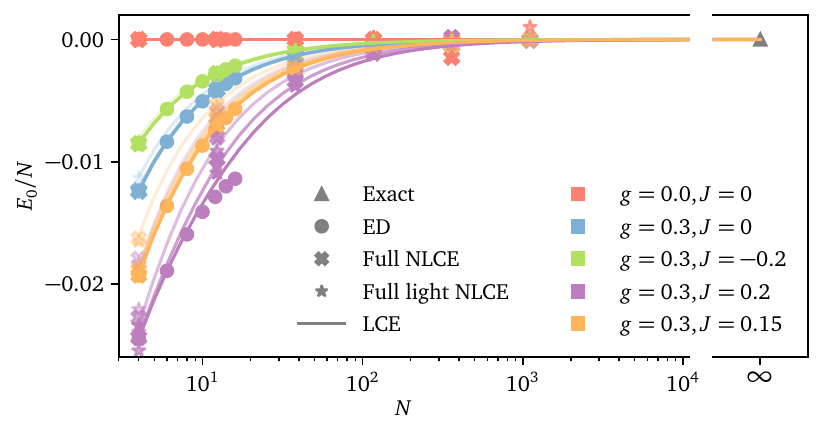}
    \caption{
        Ground-state energy density $E_0/N$ as a function of $N$ for different parameters within or close to the paramagnetic normal phase of the Dicke-Ising chain.
        The parameter points are indicated in Fig.~\ref{fig:phasediagramDI} for orientation.
        The energy density is obtained for small system sizes $N$ with ED (dots, up to $N=16$), for $N\to \infty$ with an exact theory \cite{Schellenberger2024} (triangle, only valid within the paramagnetic normal phase) and with graph expansions over the full range of $N$.
        The graph expansions are calculated perturbatively (solid lines, LCE up to perturbative order 8) and numerically (crosses for full NLCE up to order 6, stars for full light NLCE up to order 7).
        The opacity indicates the maximum order, from low (high) opacity corresponding to low (high) orders.
        The novel graph expansion methods converge to the established ED and exact theory results in the respective limits.
        As expected, the speed of convergence depends on how close the parameters are to the superradiant phase transition.
        }
    \label{fig:gs_density_mesoscopic}
\end{figure}

In Fig.~\ref{fig:gs_density_mesoscopic} we visualize the ground-state energy density for different parameters varying the number of spins $N$, which are also indicated in the phase diagram (Fig.~\ref{fig:phasediagramDI}) for orientation.
First, we take a look at the limits of microscopic and macroscopic systems.
For the case of $N\to\infty$, we can use the findings of \cite{Schellenberger2024} to map the Dicke-Ising chain to an effective Dicke model.
From that we know that the ground-state energy density is not altered by changing the parameters $g,J$, thus normalizing the density to vanish for this plot.
Be aware that this statement only holds for the normal phases of the Dicke-Ising chain, therefore making no prediction about the energy in the thermodynamic limit for data points outside of the phases \cite{Schellenberger2024}.
Next, we calculated the energy of microscopic systems with ED consisting of up to $N=16$ matter sites.
As expected, for finite systems the energy density does depend on the interaction parameters, with stronger deviations from zero for data points closer to or beyond the phase transition.

Using the novel graph expansion method for light-matter systems, we calculate the energy density over the full range of $N$ using the full matter expansion.
First, we solve the involved graphs with the perturbative method \pcst (see App.~\ref{sec:perturbative} for more information on the method) -- visualized with solid lines -- using the full matter expansion described in Subsubsec.~\ref{sec:typegraphexpansion}.
We visualize the different maximum perturbative orders of the series expansion with varying opacity to indicate the speed of convergence.
As we can treat all terms except the light-matter interaction as unperturbed Hamiltonian for the perturbation theory, we can write down the ground-state energy densities as series expansion in $g$ up to perturbative order 6 when keeping $J$ general and up to perturbative order 8 when inserting a specific value for $J$.
To be noted, we have to calculate the full matter graph expansion only up to order 3 (4) to reach perturbative order 6 (8) as every light-matter bond has to be touched by the perturbation twice, making this type of graph expansion particularly efficient.
As discussed in Subsubsec.~\ref{sec:mesoscopic}, the series expansion is given in terms of system size $N$, making the mesoscopic regime approachable.
We provide all obtained series expansions in the supplementary data \cite{Schellenberger2026_data}.

\begin{figure}
    \centering
    \includegraphics{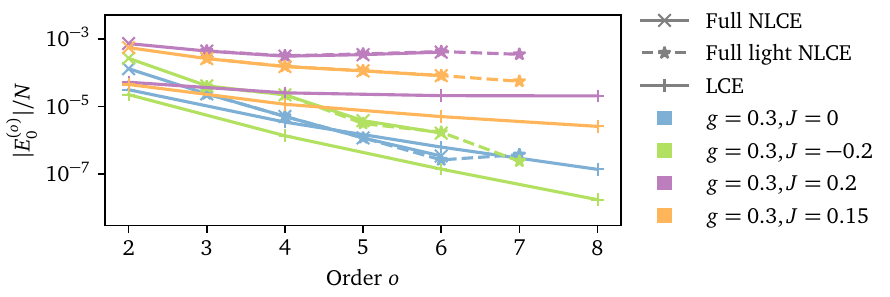}
    \caption{
        Convergence behavior of the graph-expansion methods for different parameters for the ground-state energy density for $N=60$.
        The parameter points are indicated in Fig.~\ref{fig:phasediagramDI} for orientation.
        The absolute contribution of the $o$-th order of the ground-state energy density $|E^{(o)}_0|/N$ is plotted from low to high order.
        For convergence the contributions have to decrease for larger $o$.
        Convergence is observed for all data points within the paradigmatic normal phase with different speeds.
        For the data points within the superradiant phase (purple color), no clear convergence is observed within the investigated orders.
        }
    \label{fig:convergence_gs}
\end{figure}

Additionally, we solve the graphs with ED to obtain a non-perturbative graph expansion (NLCE).
We perform both a full graph expansion and a full light expansion (see Subsubsec.~\ref{sec:typegraphexpansion} for details) up to order 6 and 7, respectively, which we will compare later in Fig.~\ref{fig:convergence_gs} in more detail.
As discussed in detail in Subsubsec.~\ref{sec:typegraphexpansion}, the number of subgraphs scales differently for the two kind of expansions, with 347 subgraphs for the full graph expansion against 66 subgraphs for the full light expansion in order 8 for the 1D case.
Higher orders suffer strongly from the high-order scaling behavior of the embedding factors, making high-precision calculations needed to obtain usable results.
We discuss this issue and possible workarounds in App.~\ref{sec:limitations}.

Comparing ED and effective theory with our results from the graph expansions, we obtain very good agreement.
As expected, when approaching the phase transition, higher orders are needed to converge to the (numerically) exact results for the limiting cases, as can be seen for the orange and purple data points.
This resembles the fact that higher orders, involving more spins, are needed to describe the dominating effects for the formation of the ground state.

To quantify the convergence of the various graph expansions, we plot the contributions from individual (perturbative) orders to the ground-state energy density in Fig.~\ref{fig:convergence_gs}.
For a graph expansion to converge well, we need the contributions to shrink for increasing order.
This indicates that the dominating processes happen at lower orders, i.e., smaller graphs.
In the figure we compare all used graph solvers, e.g., the perturbative linked-cluster expansion (LCE), the numerical full graph expansion (NLCE), and the full light graph expansion, introduced in Subsubsec.~\ref{sec:typegraphexpansion}.

In general, all three methods behave qualitatively similar.
Especially the two NLCE methods perform almost identically, justifying the more performant full light expansion for the used parameter regime.
The deviations between the LCE and NLCE methods can be explained by the different solvers, taking only perturbative or also non-perturbative processes into account.

Crucially, we obtain different speeds of convergence for the different data points, corresponding to the different negative slopes.
While we obtain clear convergence for points within the normal phase, getting slower for points close to the phase transition (orange data point), we observe no clear signs for convergence for points within the superradiant phase (purple data point), as the negative slope is vanishing.
This strongly indicates that our graph expansion approach fails outside of the normal phase, which is in line with our argumentation that the dominating effects of the ground state have to be captured mainly by small graphs.

Going to larger orders, the convergence of the NLCE method is furthermore limited by the precision of the ED calculations.
This shows up in increasing contributions at higher orders, stemming from insufficient accuracy when calculating the reduced contribution of the respective graphs (see green and blue data points).
We discuss this issue in more detail in App.~\ref{sec:limitations}.

\begin{figure}[t]
    \centering
    \includegraphics{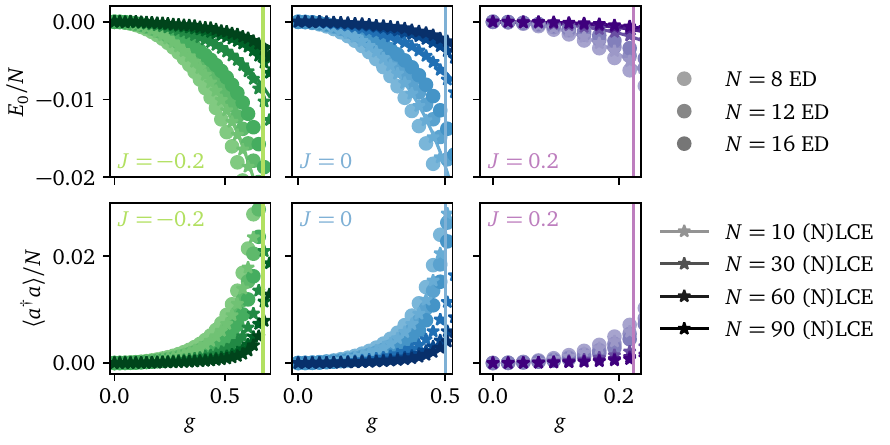}
    \caption{
        Observables of the ground state for varying light-matter couplings $g$ at different system sizes.
        We use ED for small system sizes as benchmark for the graph expansion method.
        The graph expansions are calculated perturbatively (solid lines, up to perturbative order 8) and numerically (stars, up to order 7 with full light expansion).
        The color tone indicates the system size from light to dark for small to large systems.
        Each inset fixes the Ising coupling $J$ to one of the values investigated in the other plots, too.
        The vertical line marks the second-order phase transition, as obtained from mean-field calculations \cite{Zhang2014,Schellenberger2024}.
        Upper panel: The ground-state energy density is plotted for various system sizes.
        Lower panel: The photon density of the ground state is plotted for various system sizes.
        }
    \label{fig:observables_gs}
\end{figure}

To further analyze the impact of the system size to ground-state observables, we plot the ground-state energy density and the ground-state photon density for various system sizes and Ising couplings.
For small system sizes up to $N=16$ we use again ED to verify our findings.
We know from analytical considerations \cite{Rohn_2020,Schellenberger2024} that the value of the two observables vanishes in the thermodynamic limit within the normal phase, e.g., up to the superradiant phase transition, which is indicated with solid vertical lines.
In between, we plot our graph expansion results, both perturbatively and non-perturbatively for the ground-state energy density and non-perturbatively for the photon density.
As can be seen, the graph expansion result interpolate smoothly between the two limiting cases, known from ED and the effective theory.

When comparing the behavior of the observables for different Ising couplings, we can make two observations.
First, for varying $J$ and fixed $g$ the finite-size effects are stronger (weaker) for antiferromagnetic (ferromagnetic) Ising interactions.
This can be explained by the destabilizing (stabilizing) effect of the antiferromagnetic (ferromagnetic) Ising interactions.
For antiferromagnetic Ising interactions the energy of the magnonic excitations is lowered, leading to stronger quantum fluctuations in the ground state and thus stronger finite-size effects.
Analogously, we can argue for the ferromagnetic case and the reduced finite-size effects.

Second, when moving close to the phase transition, we observe that the finite-size effects are stronger (weaker) for ferromagnetic (antiferromagnetic) Ising interactions, raising the question of the role of the Ising interaction for the critical behavior.
To investigate this in more detail, we concentrate on the series-expansion's $1/N$ contribution around the critical point for the energy density.
This contribution was investigated for the pure Dicke model ($J=0$) analytically by Vidal and Dusuel \cite{Vidal2006}.
Using the Holstein-Primakoff transformation including the $1/N$ corrections, they found the canonical form of
\begin{align}
    E_0/N = c_0 + \frac 1 N \left[ c_1 + c_2 (g_c - g)^{\theta} \right] + \mathcal O(1/N^2)
    \label{eq:dicke-critical-finite-size}
\end{align}
with constants $c_i$ depending on the system parameters and $\theta=1/2$ \cite{Vidal2006}.
This gives us an analytical starting point to check on the impact of added matter-matter interaction on the critical behavior.
In contrast to the study by Vidal and Dusuel, we can determine the $c_i$ coefficients only perturbatively.
Additionally, we have access to higher order in the $1/N$ expansion, which we will ignore in the following discussion for simplicity.

To extrapolate our extracted series beyond the convergence radius, we apply Dlog-Padé techniques.
Dlog-Padé extrapolations are a well established method to investigate critical behavior and determine critical points and exponents.
As the critical point of the superradiant phase transition is known analytically, we can also use biased Dlog Padés that use the information about the precise critical point to yield higher-precision estimates for the critical exponent.
For an in-depth introduction to these extrapolation techniques, we refer to the respective literature \cite{Adelhardt2024}.
In App.~\ref{sec:perturbative}, we present the concrete steps to derive the presented results.

\begin{figure}[t]
    \centering
    \includegraphics{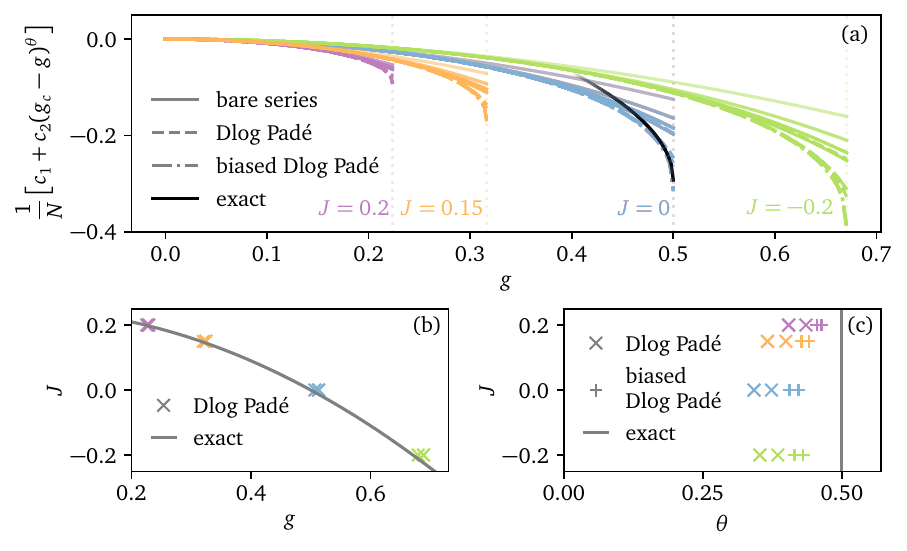}
    \caption{
        Extrapolations and bare series expansion of the ground-state energy density around the superradiant phase transition.
        The color scheme of the previous figures is used indicating the respective Ising interaction that is fixed for all plots (see Fig.~\ref{fig:phasediagramDI}).
        (a) The $1/N$ corrections of Eq.~\eqref{eq:dicke-critical-finite-size} are plotted using the bare series, dlog Padé and biased Dlog-Padé extrapolations.
        The different extrapolations are given by varying the degrees to indicate on the quality of the results.
        For the case of the Dicke model (blue color), the exact behavior around the critical point is plotted \cite{Vidal2006}.
        (b) Using the Dlog-Padé extrapolants, the critical coupling $g_c$ is extracted. 
        Multiple crosses correspond to different extrapolations.
        The solid gray line shows the exact phase boundary, from analytical considerations \cite{Zhang2014,Schellenberger2024}.
        (c) Using both biased and unbiased extrapolations, the critical exponent $\theta$ from Eq.~\eqref{eq:dicke-critical-finite-size} is plotted.
        Multiple crosses correspond to different extrapolations.
        The exact critical exponent is given in a gray solid line.
        }
    \label{fig:critical}
\end{figure}

In Fig.~\refwithlabel[(a)]{fig:critical} we plot the ground-state energy density in first order of $1/N$ up to the critical point for different Ising couplings, varying the light-matter interaction.
We plot both ordinary Dlog Padé and biased Dlog Padé alongside the bare series results.
While the extrapolations agree well with the bare-series results in the limit $g=0$, we obtain better results when moving close to the phase transition, being able to capture critical behavior.
This is shown for the limiting case of the Dicke model, where we can compare our extrapolations with the exact result from \cite{Vidal2006}.
As can be seen, we match qualitatively well with the exact result close to the phase transition using the biased Dlog-Padé extrapolations.

From the Dlog-Padé extrapolations without bias, we can get an estimate of the critical coupling \cite{Adelhardt2024}, to compare it with the known exact value \cite{Zhang2014,Schellenberger2024}.
The results are plotted in Fig.~\refwithlabel[(b)]{fig:critical} showing good agreement with the analytical line.
The small deviations can be explained by the comparably low maximum orders we have for the bare series.
Thus, it is rather surprising that the results coincide that well, potentially being grounded in the simple effective description of the model when approaching the thermodynamic limit \cite{Schellenberger2024}.
One recognizes a square-root like behavior for all plotted extrapolations, as found analytically in \cite{Vidal2006} for the $J=0$ case.
Indeed, when calculating the estimated critical exponent with both unbiased and biased extrapolations, we qualitatively find an exponent in range of $1/2$, as shown in Fig.~\refwithlabel[(c)]{fig:critical}, while systematically underestimating the expected exponent.
The exponent of $1/2$ can be confirmed to also hold for finite Ising couplings using Q-finite-size scaling for the superradiant phase transition \cite{Langheld2022,Langheld2025}.
Again, our study is mainly limited by the comparably low order of the calculated series but can confirm nonetheless the expected square root behavior.

In contrast to the critical exponent, the scaling variables $c_1, c_2$ do depend on the used Ising coupling, as can be seen by the position and the slope of the respective extrapolations as well as the bare series in Fig.~\refwithlabel[(a)]{fig:critical}.
This again raises the question on the impact the Ising coupling has on the finite-size behavior, as discussed around Fig.~\ref{fig:observables_gs}.
Rescaling both the $x$- and $y$-axis of Fig.~\refwithlabel[(a)]{fig:critical} with the respective critical-coupling value $g_c$, we find that the extrapolations lie quite close to each other (not shown).
While this is no rigorous proof, we therefore assume that the different $c_1,c_2$ values are explained by the different energy scales for the different Ising couplings.
This agrees well with the intuitive understanding of the role of the Ising coupling, being merely a rescaling factor for an effective magnetic field \cite{Schellenberger2024}.
Nonetheless, it would be interesting to investigate this feature in more detail in future studies to get an in depth explanation of the finite-size processes that happen around the phase transition.

\section{Conclusion}
\label{sec:conclusion}

In this work we presented a method to treat light-matter systems by performing a full graph expansion to obtain low-energy properties of the respective model.
We therefore extended the graph expansion technique, which is well established and successfully used for pure matter systems \cite{Domb1974, Gelfand2000, OitmaaHamerZheng2006,Knetter2000,Knetter2003_JoPA,Knetter2003}, to systems that are coupled to one or more light modes.
The structure of the graphs, which include light-matter interactions, makes it necessary to treat these graphs separately.
For doing so, we used results from non-linked cluster theory, to treat these graphs as disconnected ones \cite{Domb1974,OitmaaHamerZheng2006}.
With this approach we are not only able to calculate observables in the thermodynamic limit but also to obtain a $1/N$ series expansion to investigate the mesoscopic regime.
In contrast to many other numerical methods, we are thus able to perform calculations that are agnostic to the system size, taking the particle number as free variable.

As first application for this new approach, we investigated the Dicke-Ising chain \cite{Zhang2014} and calculated the ground-state energy density and the photon density in the normal paramagnetic phase for arbitrary system sizes.
To check for the validity of the results, we compared the graph-expansion data with ED results for small systems and an exact effective theory in the thermodynamic limit \cite{Schellenberger2024}.
We obtain good agreement between the three methods, as long as we are staying in the non-superradiant phase.
Outside of the phase, we both see that the graph expansion does not longer converge as well as the results deviate from the known results from ED and effective theory.
Furthermore, we used extrapolation techniques to extract the critical point and the critical exponent of the superradiant phase transition, by focussing on the $1/N$ contributions of the ground-state energy density \cite{Domb1989,Adelhardt2024}.

Next steps could include calculating more sophisticated observables like the entanglement entropy, correlation functions, and properties of low-lying excitations.
All these observables should be accessible with the method, analogously to the graph expansion approach for pure matter systems \cite{Gelfand2000,Trebst2000,Zheng2001,Knetter2003}.
Additionally, it would be worthwhile to investigate higher dimensions of the Dicke-Ising model to take advantage of the graph expansion being able to also tackle higher dimensions without changing the approach.

While we are able to unravel the mesoscopic regime for the Dicke-Ising chain, the results do not show any specific features unique to this regime.
Thus, it would be a major step to find models and observables that show specific features that can be captured by this method.
As there are numerous studies about stunning effect in the mesoscopic regime for closely related models \cite{Vidal2006,Lenk2022,Kudlis2023,Geng2025}, it would be a logical next step to investigate these studies in more detail.
To be able to do so, one potentially has to move to open systems or study the time evolution of low-energy states.

Apart from that, it would be valuable to apply the new method to systems, whose behavior in the thermodynamic limit is not known.
In particular, this includes systems with a matter part that is not analytically solvable, making the present approach more valuable, both for the thermodynamic limit and the mesoscopic regime.

\section*{Acknowledgements}

We thank Max Hörmann, Anja Langheld, and Jonas Leibig for fruitful discussions on the Dicke-Ising model and Anja Langheld for helping with the finite-size scaling of the Dicke-Ising model.
We further thank Matthias Mühlhauser for helping to find the fitting ansatz for the new type of light-matter graphs.
We acknowledge the use of the following python libraries to compute and visualize the presented results: \texttt{NumPy} \cite{NumPy2020}, \texttt{SciPy} \cite{SciPy2020}, \texttt{SymPy} \cite{SymPy2017}, \texttt{QuSpin} \cite{Weinberg2017}, and \texttt{Matplotlib} \cite{Matplotlib2007,Matplotlib3.6.1}.
For the \pcst calculations we use the \texttt{pcstpp\_CoefficientGenerator} \cite{Lenke2023, pcstppCoefficientGenerator1.0.0}.
For calculating the graph decomposition we use the \texttt{graph decomposition program for light-matter systems} \cite{graphdecomposition1.0.0}.

\paragraph{Author contributions}
AS: Conceptualization, Data curation, Formal analysis, Investigation, Methodology, Writing -- original draft, Writing -- review \& editing.
KPS: Conceptualization, Methodology, Supervision, Writing -- review \& editing.\footnote{Following the taxonomy \href{https://credit.niso.org}{CRediT} to categorize the contributions of the authors.}

\paragraph{Funding information}
This work was funded by the Deutsche Forschungsgemeinschaft (DFG, German Research Foundation) -- Project-ID 429529648 -- TRR 306 QuCoLiMa(Quantum Cooperativity of Light and Matter). KPS acknowledges the support by the Munich Quantum Valley, which is supported by the Bavarian state government with funds from the Hightech Agenda Bayern Plus.

\paragraph{Data availability}
Supplementary data for all figures, including the analytical results and the obtained \pcst series expansions, are available online \cite{Schellenberger2026_data}.
The program to calculate the graph expansion is available online \cite{graphdecomposition1.0.0}.

\begin{appendix}
\numberwithin{equation}{section}

\section{Perturbative calculations on the graphs}
\label{sec:perturbative}

In this appendix we will provide a short introduction into the \pcst method and ways to extrapolate the results via Padé and Dlog-Padé techniques \cite{Domb1989}.
For more details on the methods we refer to the work introducing the \pcst method \cite{Lenke2023} and to the review \cite{Adelhardt2024}, giving a general introduction to Padé and Dlog-Padé techniques.
Additionally, one finds introductions and applications to the perturbative continuous unitary transformations (pCUTs) approach -- being the foundation for the extensions of \pcst -- and to the used extrapolation techniques in the following works \cite{Domb1989,Adelhardt2020,Muehlhauser2024,Kott2024}. 

\subsection{Generalized perturbative continuous similarity transformations}
First, we provide an overview over the \pcst method.
To keep the subsection short, we focus on the aspects of the methods that are needed for this work.
One should mention that \pcst can not only be used for closed Hermitian systems -- as done here -- but can also be used for non-Hermitian and open quantum systems \cite{Lenke2023}.

The \pcst is an extension of the continuous unitary transformations (CUTs) and continuous similarity transformations (CSTs), which are well-established in the field of quantum many-body physics.
The goal of these methods is to transform the Hamiltonian $\Ham$ of the system into an effective Hamiltonian $\Ham_\mathrm{eff}$ that is easier to handle.
To do so, one defines a unitary (similarity) transformation $\mathcal S(\ell)$ when using CUTs (CSTs) that depends on the continuous parameter $\ell\in [0,\infty]$.
Thus, one writes the continuous transformation as
\begin{align}
    \mathcal H(\ell) = \mathcal S(\ell) \mathcal H \mathcal S(\ell)^{-1}
\end{align}
with $S(\ell)^{-1} = S(\ell)^{\dagger}$ when using CUTs and $\mathcal H(0) = \mathcal H$ as the starting point of the continuous transformation.
We define the sought-for effective Hamiltonian $\mathcal H_\mathrm{eff} = \mathcal H (\infty)$ as the finishing point of the continuous parameter $\ell$.

The question and main task remains to choose a suitable transformation $\mathcal S(\ell)$ to bring the Hamiltonian in a basis that is easier to handle.
For doing so, one can use the fact that $S(\ell)$ is continuous to define the attached infinitesimal generator $\eta (\ell)$ in order to write down the flow equation
\begin{align}
    \partial_\ell \mathcal H(\ell) = [\eta (\ell), \mathcal H(\ell)]\,.
\end{align}
In general these differential equations are not exactly solvable and thus have to be truncated in some fashion -- in the case of \pcst perturbatively.

This perturbative ansatz has the advantage that we can solve the transformation for larger classes of Hamiltonians, independent of a specific model.
For that we subdivide $\mathcal H = \mathcal H_0 + \mathcal V$ into an unperturbed part $\mathcal H_0$ and a perturbation $\mathcal V$, compatible with the general definition in Eq.~\eqref{eq:ham} for linked-cluster expansions.
For $\mathcal H_0$ we demand that it is a sum of equidistant ladder spectra that are bounded from below, like
\begin{align}
    \mathcal H_0 = \sum_{\alpha=1}^q \epsilon^{(\alpha)} Q^{(\alpha)}
\end{align}
with individual ladder spectra $Q^{(\alpha)}$ and individual spacings $\epsilon^{(\alpha)}$.
The perturbation $\mathcal V$ has to be decomposable into processes $T_{(m^{(1)},\dots,m^{(q)})}$ that move down and up on ladder $\alpha$ by an amount $m^{(\alpha)}\in \mathbb Z$, like
\begin{align}
    \mathcal V = \sum_{(m^{(1)},\dots,m^{(q)})} T_{(m^{(1)},\dots,m^{(q)})}\,.
\end{align}
In other words the unperturbed energy of a state is changed by
\begin{align}
    M((m^{(1)},\dots,m^{(q)})) = \sum_{\alpha=1}^q \epsilon^{(\alpha)} m^{(\alpha)}
\end{align}
when applying $T_{(m^{(1)},\dots,m^{(q)})}$.

For this type of Hamiltonian, we can solve the respective flow equations independent of the particular implementation of the model, yielding an effective Hamiltonian
\begin{align}
    \mathcal H _\mathrm{eff} = \mathcal H_0 + \sum_{k=1}^\infty \sum_{M(\bar m)= 0} C_{\bar m} \cdot T_{(m_1^{(1)},\dots,m_1^{(q)})}\cdots T_{(m_k^{(1)},\dots,m_k^{(q)})}
    \label{eq:Hameff}
\end{align}
with $\bar m$ being a combination of $k$ perturbation indices that have to fulfill
\begin{align}
    M(\bar m) := \sum_{i=1}^k M((m^{(1)}_i,\dots,m^{(q)}_i)) = 0\,.
\end{align}
Physically speaking, this ensures that $\mathcal H _\mathrm{eff}$ preserves the unperturbed energy of a state.
Solving the differential equations yields the coefficients $C_{\bar m}$ that solely depend on the spectrum of $\mathcal H_0$ and $\bar m$.
The coefficients $C_{\bar m}$ can be calculated with the program that was published on GitHub \cite{pcstppCoefficientGenerator1.0.0}.
To obtain the energy of some given state $\ket s$, one has to calculate $\braket{s | \mathcal H _\mathrm{eff} | s}$ using Eq.~\eqref{eq:Hameff}.
For this work, we use this approach to calculate the ground-state energy on graphs as explained in Sec.~\ref{sec:method}.
It is also possible to calculate other observables in the new basis by applying the continuous transformation onto the respective observable. 
See \cite{Lenke2023} for more details.

\subsection{Extrapolations of series expansions}
If one is interested in investigating quantum-critical phenomena with a perturbative series, Dlog-Padé extrapolations offer a way to extract the quantum-critical point and critical exponents.
The crucial mechanism is to rewrite the perturbative series in terms of a function that better represents the expected universal critical behavior at critical points \cite{Domb1989}.

Starting with the perturbative series of an observable $\mathcal O(\lambda)$ with perturbation parameter $\lambda$ up to the perturbative order $o_\mathrm{max}$, we define the Padé extrapolant $P[L,M]$ as
\begin{align}
    P[L,M] = \frac{P_L(\lambda)}{Q_M(\lambda)} = \frac{p_0 + p_1 \lambda + \cdots + p_L \lambda^L}{1+q_1\lambda + \cdots + q_M \lambda^M}
\end{align}
with coefficients $p_i,q_i\in \mathbb R$ and degrees $L,M$ of the polynomials in the numerator and denominator, respectively \cite{Adelhardt2024}.
We fix $L+M\leq o_\mathrm{max}$ and obtain the coefficients by a set of linear equations such that the series expansion of $P[L,M]$ is equal to $\mathcal O(\lambda)$ up to the $o_\mathrm{max}$th order.
In that sense the Padé extrapolant is again the observable in a rewritten form that can now resemble for examples divergences by having roots in the denominator.

Using the definition of Padé extrapolants, we can apply it to the logarithmic derivative of $\mathcal O(\lambda)$, namely
\begin{align}
    P[L,M] = \frac{\mathrm d}{\mathrm d \lambda} \ln (\mathcal O(\lambda))\,,
    \label{eq:dlogpade}
\end{align}
fixing $L+M\leq o_\mathrm{max}-1$ due to the applied differentiation.
To get back the observable of interest in a rewritten form, we integrate and exponentiate the result of Eq.~\eqref{eq:dlogpade}.
The form of Eq.~\eqref{eq:dlogpade} is beneficial to extract critical quantities, assuming that the observable of interest has a dominant term of the form $|\lambda -\lambda_c|^\theta$ with $\lambda_c$ being the critical point and $\theta$ the critical exponent.
Applying the logarithmic derivative on this expression, we obtain
\begin{align}
    \frac{\mathrm d}{\mathrm d \lambda} \ln \left(|\lambda -\lambda_c|^\theta\right) 
    = \frac{\theta \cdot |\lambda -\lambda_c|^{\theta-1}}{|\lambda -\lambda_c|^\theta}
    = \frac{\theta }{|\lambda -\lambda_c|}\,.
    \label{eq:criticalbehaviordlogpade}
\end{align}
Thus, we can extract $\lambda_c$ by calculating the roots of the denominator of Eq.~\eqref{eq:dlogpade} and obtain $\theta$ by differentiating the denominator of above expression at the critical point $\lambda_c$.
If the critical point $\lambda_c$ is known -- as it is the case for this work -- we can define a biased Dlog Padé by
\begin{align}
    P[L,M] = (\lambda_c - \lambda ) \frac{\mathrm d}{\mathrm d \lambda} \ln (\mathcal O(\lambda))\,.
    \label{eq:biaseddlogpade}
\end{align}
Comparing Eqs.~\eqref{eq:criticalbehaviordlogpade} and \eqref{eq:biaseddlogpade}, we obtain $\theta$ directly by calculating Eq.~\eqref{eq:biaseddlogpade} at the critical point.
It is important to state that this idealized situation is distorted by the limitations of the series expansion around a point in phase space that is not the critical point.
This may include inaccuracies due to the finite perturbative order but also avoided level crossings that limit the convergence of the series and therefore also the extrapolations.
We thus have to identify the physical and non-physical parts of the results from extrapolation to exclude any spurious extrapolants from the discussion.
We do so by checking for non-physical poles that happen way before the actual point of phase transition and close to the real axis. Specifically, for this work we identified all poles at $g/g_c< 0.8$ to be non-physical poles.
If such a pole occurs, we ignore the respective extrapolant for the discussion.
Furthermore, we only look at extrapolants with $|L-M|\leq 2$ as extrapolants with smaller $|L-M|$ are known to converge faster.

For the case of the finite-size energy corrections [see Eq.~\eqref{eq:dicke-critical-finite-size}], we do not directly have the form discussed above.
To be able to apply these extrapolation techniques, we choose the derivative of $E_0/N$ as the observable of interest.
Differentiating Eq.~\eqref{eq:dicke-critical-finite-size} gives us
\begin{align}
    \frac{\mathrm d}{\mathrm d g} \frac{E_0}{N}= -\frac {c_2\cdot \theta} N \cdot (g_c -g)^{\theta -1} + \mathcal O(1/N^2)\,.
\end{align}
As we now have a quantity that scales as assumed in the derivation above, we can apply unbiased and biased Dlog-Padé extrapolations onto this quantity.
Keep in mind, that we loose another perturbative order in our series expansion due to the additional differentiation.
In the end, we integrate the extrapolants, to obtain the ground-state energy density, which is plotted in Fig.~\ref{fig:critical}.

\section{Limitations of the non-perturbative graph expansion method}
\label{sec:limitations}

In this appendix we discuss the limitation in accuracy when using NLCE in higher orders and for larger particle numbers.
We face and mention this issue throughout this work, most prominently in Figs.~\ref{fig:gs_density_mesoscopic} and \ref{fig:convergence_gs}, where we observe diverging energies going to higher orders and larger particle numbers.

To understand this limitation better, we take a look at the general form of an intensive observable $\mathcal O(G)/N$, as introduced in Eqs.~\eqref{eq:graph-decomposition-iso} and \eqref{eq:finite-size-series-expansion} when applying the graph expansion:
\begin{align}
    \mathcal O(G) / N = \sum_{o=0}^\infty d_o N^{-o} = \sum_{\xi\subseteq C} N(\xi,C) \cdot o (\xi)\,.
    \label{eq:observable_limitation}
\end{align}
Examining the last expression, we find two parts.
First, the embedding factor $N(\xi,C)$ that we can obtain analytically as a series in $N$, as discussed in detail in Sec.~\ref{sec:method}.
Second, we calculate the reduced contribution $o (\xi)$ of a given graph $\xi$.
For the NLCE, we use exact diagonalization to calculate the respective observable $\mathcal O(\xi)$ on $\xi$ and apply Eq.~\eqref{eq:red-contr} to obtain $o (\xi)$.
Assuming the divergence to lie in numerical artifacts, we therefore have to consider the accuracy of the exact-diagonalization calculation.

Using the \texttt{quspin} package \cite{Weinberg2017}, we rely on floating point variables with double-point precision.
As an estimation of orders, we assume a precision of $16$ decimal digits, i.e., an potential absolute error of $10^{-16}$ when having parameters of order $1$ in the Hamiltonian.
To explain the divergent behavior for larger $N$, the crucial point is the scaling behavior of the embedding factor $N(\xi,C)$.
For the discussed model, a graph of order $n$ (independent of whether it is a full graph expansion or a full light expansion) has an embedding factor up to order $n$ in $N$.
Only taking this highest order into account, in the general case we have an embedding factor in the order of $10^{e \cdot n}$ for $N=10^{e}$, leading to an absolute error in the order of $10^{e \cdot n- 16}$.
With our investigated maximum order of $n=7$, we therefore end up with a potential error in the realm of $10^{-2}$ ($10^{5}$) for $N=100$ ($N=1000$).
This explains why we expect to have diverging errors for higher orders $n$ and larger particle numbers $N$.
It is to be noted that the problem can not be circumvented by only looking at lower orders of the series expansion in $N$ as written in Eq.~\eqref{eq:observable_limitation} because the different orders in $N$ mix between the embedding factor and the reduced contribution, i.e., also larger orders of the embedding factor contribute to the lower orders in $N$ of a respective observable.

To deal with this problem in general, we need to calculate $o (\xi)$ exactly or with more precision.
While the first is not possible in general, for the latter, one would have to perform calculations with data types allowing for higher precision, which is not possible with the \texttt{quspin} package, up to our knowledge.
The drawbacks of this approach are higher computation times, due to the higher precision, and the strong expected scaling of the absolute error, making very high precision needed when going to higher order and larger particle numbers.

These limitations do not apply for the perturbative graph solver.
The reason for this is the possibility to keep $N$ as a free parameter throughout the whole calculation.
Therefore the quantities of very different orders of magnitude do not mix, as happening for the NLCE approach.
\end{appendix}



\bibliography{bibliography.bib}


\end{document}